\documentclass[showpacs,preprintnumbers,amsmath,amssymb,aps,prd,nofootinbib,unicode-math]{revtex4-2}
\pdfoutput=1

\usepackage{graphicx}
\usepackage{amsmath,amssymb}
\usepackage{epstopdf}
\graphicspath{{./Figures/}}

\usepackage{color}%

\usepackage[colorlinks,urlcolor=blue,citecolor=blue]{hyperref}
\def\hhref#1{\href{http://arxiv.org/abs/#1}{arXiv:#1}} 

\begin{document}

\title{Heisenberg-Euler and the Quantum Dilogarithm}

\author{Gerald V. Dunne}
\affiliation{Physics Department, University  of Connecticut, Storrs, CT 06269}

\begin{abstract}

A dispersion integral representation of the Heisenberg-Euler QED effective lagrangian is derived, with Faddeev's quantum dilogarithm as a generalized Borel kernel. The nonperturbative imaginary part of the effective lagrangian is expressed as the quantum dilogarithm, while the real part has the form of a dispersion integral involving both the quantum dilogarithm and its modular dual, a manifestation of electromagnetic duality. The Heisenberg-Euler effective lagrangian generates all one-loop QED scattering amplitudes in a constant external field, with the Lorentz invariants of the constant background electromagnetic field playing the role of the Mandelstam variables in conventional QED dispersion theory.

\end{abstract}


\maketitle

\section{Introduction}
\label{sec:intro}

The one-loop QED effective lagrangian encodes the nonlinear and nonperturbative physics of the effective dynamics of the photon field, after integrating out the massive electron and positron fields 
\cite{euler,weisskopf,nikishov,dittrich-reuter,Dittrich:2000zu,dunne-kogan,Schwartz:2014sze}. It was first computed, in the low-energy limit of electrons and positrons in a constant electromagnetic field, by Heisenberg and Euler \cite{euler} in
a prescient paper very much in the modern spirit of effective field theory.
It was later formalized for more general background fields by Schwinger, Feynman, Nambu and Morette
\cite{Schwinger:1948yk,schwinger,Feynman:1949zx,Feynman:1950ir,nambu,Morette:1951zz}. 
Heisenberg and Euler expressed their result in the form of a Borel integral, which  can be written as
\begin{equation}
    \mathcal L(B, E)=-\frac{1}{8\pi^2} \int_0^\infty ds\, e^{-m^2 s}
    \left[\frac{1}{s^3}\left((e B s)(e E s)\coth(e B s) \cot(e E s)-1+\frac{e^2 s^2}{3}(E^2-B^2)\right)\right]
    \label{eq:l1be-sp0}
\end{equation}
Here the (constant) background electromagnetic field has both a magnetic and an electric component, chosen to be parallel (this is always possible for $\vec{B}\cdot\vec{E}\neq 0$), and we choose $e B, e E\geq 0$ \cite{euler,dunne-kogan}, where $e$ is the electron charge. The electron mass is $m$. The expression \eqref{eq:l1be-sp0} is still formal because of the Borel poles on the integration contour at $s_k=k \pi/(e E)$, for $k=1, 2, 3, \dots$. These poles produce an imaginary part in the effective lagrangian, which is interpreted physically as the instability of the QED vacuum under the influence of an electric field \cite{euler,schwinger,nikishov,dunne-kogan}.\footnote{This is analogous to the phenomenon of atomic ionization under the influence of an electric field \cite{silverstone}.} For example, for a purely electric constant background field Schwinger expressed the result as a real Cauchy principal value (${\mathcal P\mathcal V}$) integral and an imaginary part expressed as a sum over nonperturbative terms \cite{schwinger}:
\begin{equation}
    \mathcal L(E)=-\frac{1}{8\pi^2} {\mathcal P\mathcal V}\int_0^\infty ds\, e^{-m^2 s}
    \left[\frac{1}{s^3}\left((e E  s)  \cot(e E s)-1+\frac{(e E  s)^2}{3}\right)\right]
    + i \frac{e^2 E^2}{8\pi^3}\sum_{k=1}^\infty \frac{1}{k^2} \, e^{-m^2\pi k/(e E)} 
    \label{eq:l1e-sp0}
\end{equation}
The imaginary part of the Heisenberg-Euler (HE) effective lagrangian yields the rate of electron-positron pair production from the QED vacuum, due to the application of an external electric field, commonly known as the ``Schwinger effect''. While this effect is too weak to have been directly observed, related QED processes approaching this ultra-high intensity regime are the subject of current experimental  proposals \cite{Yakimenko:2018kih,luxe,facet,Sarri:2025qng,dipiazza}, and corresponding recent theoretical progress in high-intensity QFT
\cite{DiPiazza:2011tq,Gonoskov:2021hwf,Fedotov:2022ely}.

In general the one-loop QED effective action is given by \cite{euler,weisskopf,nikishov,dittrich-reuter,Dittrich:2000zu,dunne-kogan,Schwartz:2014sze,Schwinger:1948yk,schwinger,Feynman:1949zx}
\begin{eqnarray}
\mathcal S=-i\ln \det (i D \hskip -7pt \slash-(m-i\epsilon))
\label{eq:he-action}
\end{eqnarray}
where the Dirac operator is $D \hskip -7pt \slash =\gamma^\nu\left(\partial_\nu+ie A_\nu\right)$, and $A_\nu$ is a fixed classical gauge potential with field strength tensor $F_{\mu\nu}=\partial_\mu A_\nu- \partial_\nu A_\mu$. The effective action $\mathcal S$ is a transform of the trace of the Green's function, and the Feynman $i\epsilon$ prescription, $m\to m-i\epsilon$, expresses causality for QED with both electrons and positrons \cite{Feynman:1949zx,Feynman:1950ir,nambu,Schwinger:1948yk,schwinger}. Therefore we expect the effective lagrangian to satisfy a dispersion relation, relating the real and imaginary parts. This fundamental property is not immediately manifest in the standard HE expressions \eqref{eq:l1be-sp0}-\eqref{eq:l1e-sp0}. However, for the electric background a simple transformation (see section \ref{sec:classical}) in Eq \eqref{eq:l1e-sp0} shifts the sum over Borel poles into a modified Borel kernel, thereby re-expressing ${\mathcal L}(E)$ in a form in which the connection between real and imaginary parts is more apparent:
\begin{equation}
\mathcal{L}\left(E\right)=  \frac{e^2 E^2}{8\pi^4}\, \left\{ 2\, {\mathcal P\mathcal V} \int_0^\infty d\omega \,{\rm Li}_2\left(e^{-\omega}\right)\left[ \frac{\omega}{\left(\frac{\pi m^2}{e E}\right)^2-\omega^2} \right] + i\,   \pi \, {\rm Li}_2\left(e^{-\pi \, m^2/(e E)}\right)\right\}
\label{eq:l1e-dilog0}
\end{equation}
Here ${\rm Li}_2(x)$ is the {\it classical} dilogarithm function:
\begin{eqnarray}
    {\rm Li}_2(x)=\sum_{k=1}^\infty \frac{x^k}{k^2}
    \label{eq_li2}
\end{eqnarray}
The representation \eqref{eq:l1e-dilog0} makes it clear that the real part of ${\mathcal L}(E)$ is an integral transform of the imaginary part.

The main part of the paper, in sections \ref{sec:quantum-compact} and \ref{sec:quantum-non-compact}, is devoted to finding a dispersive representation of $\mathcal{L}\left(B, E\right)$ in the general constant background field with {\it both} an electric and magnetic component. The existence of such a dispersion representation was in fact suggested long ago by Lebedev and Ritus \cite{lebedev}. Here we show how this idea is explicitly realized for the Heisenberg-Euler effective lagrangian. This leads to several new integral representations of the HE effective lagrangian, in which the classical dilogarithm in \eqref{eq:l1e-dilog0} is generalized to the {\it quantum} dilogarithm function, a function with many interesting properties and applications in physics and mathematics \cite{faddeev,kirillov,zagier,fock,Kashaev:2011se,Garoufalidis:2021nwy}.  Here we argue that the quantum dilogarithm is naturally related to the HE effective lagrangian, and conversely some of the properties of the quantum dilogarithm have direct physical interpretation in QED. For example, electromagnetic duality \cite{Witten:1995gf} is reflected in the appearance of the quantum dilogarithm and its modular dual, and the relations between spinor and scalar QED are encoded in  functional relations satisfied by the quantum dilogarithm. See sections \ref{sec:quantum-compact} and \ref{sec:quantum-non-compact}.

The primary motivation for this analysis is to recast the familiar dispersive methods for QFT amplitudes \cite{bj} in terms of the {\it generator} of amplitudes with any number of external legs, rather than for individual diagrams with a fixed number of external legs. This generator is the effective action, and for one-loop scattering amplitudes the simplest such generator is the HE effective action. 
Already in this example we find interesting mathematical structure, for which the QED perspective provides a natural physical basis. A motivating analogy is the following: scattering amplitudes for 2-particle to 2-particle scattering are functions of two independent Lorentz invariant Mandelstam (${\rm momentum}^2$) variables, and the analytic continuation with respect to these two variables  connects different scattering channels. 
The HE effective lagrangian is also described by two independent Lorentz variables, $(\vec{E}^2-\vec{B}^2)$ and $\vec{B}\cdot\vec{E}$, and the analytic continuation structure relates perturbative and nonperturbative amplitudes.

\section{Heisenberg-Euler Effective Lagrangian and the Classical Dilogarithm}
\label{sec:classical}

As mentioned in the introduction, there is a simple connection between the {\it classical} dilogarithm function and the HE effective Lagrangian, for either a constant background magnetic field $B$ or a constant background electric field $E$. For a constant magnetic field $B$, the general Borel integral representation \eqref{eq:l1be-sp0} reduces to
\begin{equation}
\mathcal{L}\left(B\right)=
- \frac{1}{8\pi^2}\int_0^\infty 
ds\, e^{-m^2 s}\, \left[\frac{1}{s^3}\left((e B s)\coth (e B  s)-1-\frac{(e B s)^2}{3}\right) \right]
\label{eq:l1b}
\end{equation}
The subtraction terms in \eqref{eq:l1b} are the zero-field subtraction and the (on-shell) charge renormalization subtraction, respectively \cite{euler,weisskopf,nikishov,dittrich-reuter,Dittrich:2000zu,dunne-kogan,Schwartz:2014sze}. 
The Borel transform is meromorphic, with simple poles on the imaginary axis at $s=i\,k\,\pi/(e B)$,  $k\in \mathbb Z\setminus\{0\}$:
\begin{eqnarray}
    \frac{1}{s^3}\left((e B s)\coth (e B s)-1-\frac{(e B s)^2}{3}\right)=-\frac{2(e B)^2}{\pi^2}\sum_{k=1}^\infty \frac{s}{k^2 \left(\left(\frac{k \pi}{e B}\right)^2+s^2\right)} 
\label{eq:id1}
\end{eqnarray}
Rescaling $s\to k s$, we  transfer the sum to the exponential factor in the integrand of \eqref{eq:l1b}, leading  to a modified integral representation:
\begin{equation}
\mathcal{L}\left(B\right)=
\frac{e^2 B^2}{4\pi^4}\int_0^\infty d{s} {\rm Li}_2\left(e^{- m^2\, s}\right) \left[\frac{s}{\left(\frac{\pi}{e B}\right)^2+s^2}\right]
\label{eq:l1b-dilog}
\end{equation}
The expression \eqref{eq:l1b-dilog} can be interpreted as a modified "dilog-Borel" integral, with the classical dilogarithm function as the weight factor, and a corresponding modified Borel transform function: $\frac{s}{\left(\pi/(e B)\right)^2+s^2}$. The infinite tower of complex instantons (singularities of the Borel transform in \eqref{eq:id1}) has been summed, leaving just a single complex conjugate pair of poles at $s=\pm i \pi/(e B)$. The weak field ($e B\ll m^2$) expansion is generated by the modified moments:
\begin{eqnarray}
    \int_0^\infty ds\, e^{-m^2 s}\, s^{2n+1}= \frac{\Gamma(2n+2)}{m^{4n+4}} 
    \quad \rightarrow \quad  
   \int_0^\infty ds\, {\rm Li}_2\left(e^{-m^2 s}\right)\, s^{2n+1}= \frac{\Gamma(2n+2)\zeta(2n+4)}{m^{4n+4}}
    \label{eq:dilog-moments}
\end{eqnarray}
Therefore the weak field expansion of \eqref{eq:l1b-dilog} produces the familiar asymptotic series with sign-alternating factorially divergent coefficients, multiplied by a zeta function factor which encodes the instanton sum \cite{dunne-kogan}.
The strong field ($e B\gg m^2$) expansion follows from the large $s$ behavior of the modified Borel transform, $\frac{s}{\left(\pi/(e B)\right)^2+s^2}$, producing the characteristic logarithmic strong field behavior, 
\begin{eqnarray}
    \mathcal L(B)\sim -\frac{e^2}{4\pi}\, \beta_1 \mathcal L_{\rm Maxwell}(B)\ln(2 e B/m^2) \qquad , \quad e B\gg m^2 
    \label{eq:l1b-log}
\end{eqnarray}
The fine structure constant is $\alpha=\frac{e^2}{4\pi}$ in our chosen units, the Maxwell lagrangian is  ${\mathcal L}_{\rm Maxwell}(B)=-\frac{1}{2} B^2$, and 
$\beta_1=\frac{1}{3\pi}$ is the leading QED beta function coefficient, which follows since ${\rm Li}_2(1)=\zeta(2)=\frac{\pi^2}{6}$.

Integrating by parts, expression  \eqref{eq:l1b-dilog} can also be written as
\begin{equation}
\mathcal{L}\left(B\right)=
\frac{e^2 B^2}{4\pi^4}\int_0^\infty d{s} \frac{1}{e^{m^2\, s}-1} \left[-s+\frac{i \pi}{2 e B} \left(1-i \frac{e B s}{\pi}\right) \log (1-i \frac{e B s}{\pi})-\frac{i \pi}{2 e B} \left(1+i \frac{e B s}{\pi}\right) \log (1+i \frac{e B s}{\pi})\right]
\label{eq:l1b-bernoulli}
\end{equation}
This is an alternative resummation of the Borel poles, yielding another type of modified Borel kernel and transform, which has been discussed in a different context in \cite{Hatsuda:2015owa}. The integrands in \eqref{eq:l1b}, \eqref{eq:l1b-dilog} and \eqref{eq:l1b-bernoulli} have similar shape along the positive real $s$ axis. However, the modified 
``Borel transform'' in \eqref{eq:l1b-dilog} has the simplest analytic structure.
More importantly, the dispersive nature of the HE lagrangian is explicitly manifest when we use \eqref{eq:l1b-dilog} to analytically continue from a purely magnetic background to a purely electric background, $B^2 \to - E^2$, yielding the integral representation
\begin{equation}
\mathcal{L}\left(E\right)=  \frac{e^2 E^2}{8\pi^4}\, \left\{ 2\, {\mathcal P\mathcal V} \int_0^\infty ds \,{\rm Li}_2\left(e^{-m^2 s}\right)\left[ \frac{s}{\left(\frac{\pi}{e E}\right)^2-s^2} \right] + i\,   \pi \, {\rm Li}_2\left(e^{-\pi \, m^2/(e E)}\right)\right\}
\label{eq:l1e-dilog}
\end{equation}
which is the dispersive expression \eqref{eq:l1e-dilog0}.
The imaginary part is the classical dilogarithm, and the real part is a principal value integral transform of this imaginary part. 
The real part in \eqref{eq:l1e-dilog}  can alternatively be expressed as a convergent instanton sum over exponential integral Ei functions, in agreement with known results \cite{valluri}:
\begin{equation}
{\rm Re}\left[\mathcal{L}\left(E\right)\right] =
\frac{e^2 E^2}{8\pi^4} \sum_{k=1}^\infty \frac{1}{k^2} \left(e^{k\pi m^2/(e E)}{\rm Ei}\left(-\frac{k\pi m^2}{E}\right)+e^{-k\pi m^2/(e E)}{\rm Ei}\left(\frac{k\pi m^2}{E}\right)\right)
\label{eq:l1e-real}
\end{equation}

An analogous resummation for scalar QED \cite{weisskopf,dunne-kogan} in a magnetic background yields 
\begin{eqnarray}
\mathcal{L}_{\rm scalar}\left(B\right)
&=&
\frac{1}{2}\cdot \frac{e^2 B^2}{8\pi^2}
\int_0^\infty ds\,  
e^{-m^2 s} \left[\frac{1}{s^3}\left(\frac{e B s}{\sinh (e B  s)}-1+\frac{(e B s)^2}{6}\right)\right]
\label{eq:scalar-trig}\\
&=& 
-\frac{1}{2}\cdot \frac{e^2 B^2 }{4\pi^4}\int_0^\infty d{s} \, {\rm Li}_2\left(-e^{-m^2\, s}\right) \left[\frac{s}{\left(\frac{\pi}{e B}\right)^2+s^2}\right]
\label{eq:l1b-sc}
\end{eqnarray}
In the dilog form, the only differences between the scalar QED result \eqref{eq:l1b-sc} and the spinor QED result \eqref{eq:l1b-dilog} are the overall spin factor, and the minus sign in the argument of the dilogarithm, exactly as expected for bosons versus fermions.\footnote{The Fermi-Dirac and Bose-Einstein distributions are naturally associated with the (classical) polylogarithm functions \cite{dingle, pathria}.} The weak field and strong field expansions follow as in the spinor case.
The strong field limit produces the physical logarithmic behavior, with the scalar QED beta function coefficient, $\beta_1=\frac{1}{12\pi}$, arising from ${\rm Li}_2(-1)=-\pi^2/12$:
\begin{eqnarray}
    \mathcal L_{\rm scalar}(B)\sim -\frac{e^2}{4\pi} \, \beta_1 \mathcal L_{\rm Maxwell}(B)\ln(2 e B/m^2) \qquad , \quad e B\gg m^2 
    \label{eq:l1scb-log}
\end{eqnarray}
The analytic continuation from a magnetic to electric background field yields a dispersive representation:
\begin{equation}
\mathcal{L}_{\rm scalar}\left(E\right)=
 -\frac{e^2 E^2}{8\pi^4} {\mathcal P \mathcal V} \int_0^\infty d s \,{\rm Li}_2\left(-e^{- m^2\, s}\right)\left[ \frac{s}{\left(\frac{\pi}{e E}\right)^2-s^2} \right] - i\,  \frac{e^2 E^2}{16\pi^3}\, {\rm Li}_2\left(-e^{-\pi \, m^2/(e E)}\right)
\label{eq:l1sce-dilog}
\end{equation}
Once again, this expression encodes the correct analytic properties of the scalar QED effective lagrangian. 
The principal value real part can be expressed as a convergent sum, analogous to \eqref{eq:l1e-real}:
\begin{equation}
{\rm Re}\left[\mathcal{L}_{\rm scalar}\left(E\right)\right] =
-\frac{e^2 E^2}{16\pi^4} \sum_{k=1}^\infty \frac{(-1)^{k}}{k^2} \left(e^{k\pi m^2/(e E)}{\rm Ei}\left(-\frac{k\pi m^2}{e E}\right)+e^{-k\pi m^2/(e E)}{\rm Ei}\left(\frac{k\pi m^2}{e E}\right)\right)
\label{eq:l1e}
\end{equation}

We also note that the classical dilogarithm identity, ${\rm Li}_2(x)+{\rm Li}_2(-x)=\frac{1}{2} {\rm Li}_2(x^2)$, implies the scaling relations:
\begin{eqnarray}
    \mathcal L_{\rm scalar}(B)=\frac{1}{2} \mathcal L_{\rm spinor}(B)-\mathcal L_{\rm spinor}(B/2)
    \qquad; \qquad
    \mathcal L_{\rm scalar}(E)=\frac{1}{2} \mathcal L_{\rm spinor}(E)-\mathcal L_{\rm spinor}(E/2)
    \label{eq:dilog-id1}
\end{eqnarray}
Inverting \eqref{eq:dilog-id1}, the spinor QED HE effective lagrangian can be expressed as an infinite sum of  scalar QED HE effective lagrangians:\begin{eqnarray}
    \mathcal L_{\rm spinor}(B)=\sum_{k=1}^\infty 2^k \mathcal L_{\rm scalar}(2^{1-k}\, B)
    \qquad; \qquad\mathcal L_{\rm spinor}(E)=\sum_{k=1}^\infty 2^k \mathcal L_{\rm scalar}(2^{1-k}\, E)
\end{eqnarray}
For an electric background this scaling applies to both the real and imaginary parts. These scaling relations also follow from the elementary trigonometric identities
\begin{eqnarray}
    \frac{1}{\sinh(s)}=-\coth(s)+\coth(s/2)
    \qquad; \qquad 
    \frac{1}{\sin(s)}=-\cot(s)+\cot(s/2)
    \label{eq:classical-identity}
\end{eqnarray}
These scaling relations generalize when the constant background field has {\it both} magnetic and electric components: see Eqs \eqref{eq:sp-sc}-\eqref{eq:quantum-identity}.

\section{Heisenberg-Euler Effective Lagrangian and the Compact Quantum Dilogarithm}
\label{sec:quantum-compact}

Now consider the spinor QED Heisenberg-Euler effective lagrangian with {\it both} a magnetic and an electric background field, chosen to be parallel (always possible for $\vec{B}\cdot\vec{E}\neq 0$), and recall that we choose $e B, e E\geq 0$ \cite{euler,dunne-kogan}:
\begin{equation}
    \mathcal L(B, E)=-\frac{1}{8\pi^2} \int_0^\infty ds\, e^{-m^2 s}
    \left[\frac{1}{s^3}\left((e B s)(e E s)\coth(e B s) \cot(e E s)-1+\frac{e^2 s^2}{3}(E^2-B^2)\right)\right]
    \label{eq:l1be-sp}
\end{equation}
The identity \eqref{eq:id1} generalizes to the partial fraction identity \cite{ramanujan, valluri} 
\begin{eqnarray}
  \frac{1}{s^3}\left((e B s)(e E s)\coth(e B s) \cot(e E s)-1 +\frac{e^2 s^2}{3}(E^2-B^2)\right) &=& 
  -\frac{2 e^2 B E}{\pi} \sum_{k=1}^\infty \frac{\coth\left(\frac{E}{B}\, k \pi\right) }{k} \frac{s}{\left(\frac{k \pi}{e B}\right)^2+ s^2} 
  \nonumber\\
  && 
  -\frac{2 e^2 B E}{\pi} \sum_{k=1}^\infty \frac{\coth\left(\frac{B}{E}\, k \pi\right)}{k} \frac{s}{\left(\frac{k \pi}{e E}\right)^2- s^2} 
  \label{eq:id2}
\end{eqnarray}
Rescaling $s\to k\, s$, the Borel integral \eqref{eq:l1be-sp} can be expressed as a modified Borel integral:
\begin{eqnarray}
  \mathcal L(B, E) &=& \frac{e^2 B E}{2\pi^3} 
  \int_0^\infty ds \left(\left\{ {\rm Li}_2\left(e^{-m^2 s}; e^{-2\pi \frac{E}{B}}\right)+
  \frac{1}{2} \ln\left(1-e^{-m^2 s}\right) \right\}
  \left[\frac{s}{\left(\frac{\pi}{e B}\right)^2+ s^2}\right] \right.
  \nonumber\\
  && \hskip 2cm \left.
  +\left\{ {\rm Li}_2\left(e^{- m^2 s}; e^{-2\pi \frac{B}{E}}\right)+
  \frac{1}{2} \ln\left(1-e^{- m^2 s}\right) \right\} \left[ \frac{s}{\left(\frac{\pi}{e E}\right)^2- s^2} \right]\right)
  \label{eq:l1beq-sp1}
  \end{eqnarray}
Here ${\rm Li}_2(a; q)$ is the (compact) {\it quantum} dilogarithm function \cite{faddeev,kirillov,zagier,fock,Kashaev:2011se,Garoufalidis:2021nwy}:
\begin{eqnarray}
    {\rm Li}_2(a; q) =\sum_{k=1}^\infty \frac{1}{k} \frac{a^k}{(1-q^k)} 
    \label{eq:li2}
\end{eqnarray}
In the first line of \eqref{eq:l1beq-sp1} we have 
\begin{eqnarray}
   q = e^{-2\pi \frac{E}{B}} \qquad &{\rm and}& \qquad
   a= e^{-m^2 s}
   \label{eq:q1}
\end{eqnarray}
while in the second line of \eqref{eq:l1beq-sp1} we have
\begin{eqnarray}
   q = e^{-2\pi \frac{B}{E}} \qquad &{\rm and}& \qquad
   a= e^{-m^2 s}
   \label{eq:q2}
\end{eqnarray}
As before, the expression \eqref{eq:l1beq-sp1} is formal because of the pole on the contour of integration at $s=\pi/(e E)$. More precisely, the HE lagrangian $\mathcal L(B, E)$ has both a real part (the first two lines) and an imaginary part (the third line):
\begin{eqnarray}
  \mathcal L(B, E) &=& \frac{e^2 B E}{2\pi^3} 
  \int_0^\infty ds \left\{ {\rm Li}_2\left(e^{-m^2 s}; e^{-2\pi \frac{E}{B}}\right)+
  \frac{1}{2} \ln\left(1-e^{- m^2 s}\right) \right\}
 \left[\frac{s}{\left(\frac{\pi}{e B}\right)^2+ s^2}\right]
  \nonumber\\
  && 
  + \frac{e^2 B E}{2\pi^3} {\mathcal P \mathcal V}
  \int_0^\infty ds\left\{ {\rm Li}_2\left(e^{-m^2 s}; e^{-2\pi \frac{B}{E}}\right)+
  \frac{1}{2} \ln\left(1-e^{-m^2 s}\right) \right\} 
  \left[ \frac{s}{\left(\frac{\pi}{e E}\right)^2- s^2} \right]
  \nonumber\\
  &&+ i \frac{e^2 B E}{4\pi^2}\left\{ {\rm Li}_2\left(e^{-\pi m^2/(e E)}; e^{-2\pi \frac{B}{E}}\right)+\frac{1}{2} \ln\left(1-e^{-\pi m^2/(e E)}\right)\right\}
  \label{eq:l1beq-sp11}
\end{eqnarray}
The expression \eqref{eq:l1beq-sp11} has an elegant physical interpretation, using Nikishov's ``virial" representation of the imaginary part of the effective lagrangian \cite{nikishov}
\begin{eqnarray}
    {\rm Im}\left[ \mathcal L(B, E)\right]=-\frac{1}{2} {\rm tr} \ln\left(1-\langle N\rangle \right)
    \label{eq:virial1}
\end{eqnarray}
Here $\langle N\rangle$ is the average number of electron-positron pairs produced by the background electric field in a given eigenmode of the Dirac operator:
\begin{eqnarray}
    \langle N\rangle = \exp\left(-\frac{\pi}{e E}\left(m^2 + e B(2n+1\pm 1)\right)\right)
    \qquad; \quad n=0, 1, 2, ...
    \label{eq:average}
\end{eqnarray}
The $\pm$ refers to the two spin states, and there is a  Landau degeneracy factor $\left(\frac{e B}{2\pi}\right)\left(\frac{e E}{2\pi}\right)$, so the trace yields
\begin{eqnarray}
    {\rm Im}\left[ \mathcal L(B, E)\right]
    =\frac{1}{2} \left(\frac{e B}{2\pi}\right)\left(\frac{e E}{2\pi}\right) 
    \sum_{k=1}^\infty \frac{1}{k} \left[-\exp\left(-\frac{ \pi\, m^2\, k}{e E}\right)+2\sum_{n=0}^\infty \exp\left(-\frac{\pi\, (m^2+2 B n)\, k}{e E}\right)
    \right]
    \label{eq:virial2}
\end{eqnarray}
So the lowest Landau level  $(n=0)$  produces the log term in the imaginary part ${\rm Im}\left[ \mathcal L(B, E)\right]$ in \eqref{eq:l1beq-sp11}, while the higher Landau levels produce the quantum dilogarithm term.

From the ``classical limit'' ($\hbar\to 0$) of the quantum dilogarithm \cite{faddeev,kirillov,zagier,fock,Kashaev:2011se,Garoufalidis:2021nwy}
\begin{eqnarray}
    {\rm Li}_2\left(e^{-u}; e^{-\hbar}\right) \sim \frac{1}{\hbar} \, {\rm Li}_2\left(e^{-u}\right)-\frac{1}{2} \ln\left(1-e^{-u}\right)+\dots
    \qquad, \quad \hbar\to 0^+
    \label{eq:classical}
\end{eqnarray}
the general expression \eqref{eq:l1beq-sp11} reduces to the magnetic field result \eqref{eq:l1b-dilog} in the limit $E\to0$, and to the electric field result \eqref{eq:l1e-dilog} in the limit  $B\to 0$.

The $q$ parameters in \eqref{eq:q1}-\eqref{eq:q2} are related by the modular S-duality transformation \cite{Witten:1995gf,valluri,Gang:2017hbs}: 
\begin{equation}
    \hbar := 2\pi \frac{B}{E}\to \frac{4\pi^2}{\hbar}= 2\pi\frac{E}{B}
    \label{eq:modular}
\end{equation}
This means that the integral representation \eqref{eq:l1beq-sp11} is a dispersive representation in which the real part is expressed as an integral transform of the imaginary part and also of its modular dual. This is a consequence of electromagnetic duality. And the quantum dilogarithm expression \eqref{eq:l1beq-sp11} is an explicit realization of an old suggestion of Lebedev and Ritus to express the Heisenberg-Euler effective lagrangian as a dispersion integral \cite{lebedev}. See also equations \eqref{eq:scalar-new} and \eqref{eq:spinor-new} below.

 Note that the quantum dilogarithm sums in \eqref{eq:li2} converge because in the physical context of the HE effective lagrangian we have $e B, eE \geq 0$, and so $|e^{-2\pi \frac{B}{E}}|<1$ and $|e^{-2\pi \frac{E}{B}}|<1$. In this situation it is convenient to rewrite the quantum dilogarithm factors in the integrand of \eqref{eq:l1beq-sp1} in terms of the $q$-Pochhammer function:
 \begin{eqnarray}
    {\rm Li}_2(a; q) 
    = -\ln\left[ (a; q)_\infty\right]
    \label{eq:li2b}
\end{eqnarray}
Here $(a; q)_\infty$ is the $q$-Pochhammer function  \cite{qpochhammer}:
\begin{eqnarray}
    (a; q)_\infty := \prod_{r=0}^\infty (1-a\, q^r)
    \label{eq:qpoch}
\end{eqnarray}
Using the identity $(a; q)_\infty=(1-a)(q\, a; q)_\infty$, we can therefore write the HE effective lagrangian in \eqref{eq:l1beq-sp1} as
\begin{eqnarray}
  \mathcal L(B, E) 
  &=& -\frac{e^2 B E}{4\pi^3} 
  \int_0^\infty ds 
  \ln\left\{
  \left(e^{- m^2 s}; e^{-2\pi \frac{E}{B}}\right)_\infty 
  \left(e^{-2\pi \frac{E}{B}}\, e^{- m^2 s}; e^{-2\pi \frac{E}{B}}\right)_\infty
  \right\}
\left[\frac{s}{\left(\frac{\pi}{e B}\right)^2+ s^2}\right]
  \nonumber\\
  && -\frac{e^2 B E}{4\pi^3} {\mathcal P\mathcal V}
  \int_0^\infty ds
  \ln\left\{
  \left(e^{- m^2 s}; e^{-2\pi \frac{B}{E}}\right)_\infty 
  \left(e^{-2\pi \frac{B}{E}}\, e^{- m^2 s}; e^{-2\pi \frac{B}{E}}\right)_\infty
  \right\}
  \left[\frac{s}{\left(\frac{\pi}{e E}\right)^2- s^2}\right]
  \nonumber\\
  &&-i\, \frac{e^2 B E}{8\pi^2} \ln \left[\left(e^{-\pi m^2/(e E)}; e^{-2\pi \frac{B}{E}}\right)_\infty 
    \left(e^{-2\pi \frac{B}{E}}\, e^{-\pi m^2/(e E)}; e^{-2\pi \frac{B}{E}}\right)_\infty
    \right] 
  \label{eq:l1beq-sp2}
\end{eqnarray}
  
The quantum dilogarithm expressions \eqref{eq:l1beq-sp11} and \eqref{eq:l1beq-sp2} naturally split into three parts. The first two lines are both real: the first line is manifestly real, like the magnetic field case in \eqref{eq:l1b-dilog}, and the second line is a real principal value integral, like the electric field case in \eqref{eq:l1e-dilog}. The third line is the imaginary part, coming from the pole at $s=\pi/(e E)$:
\begin{eqnarray}
    {\rm Im}\left[\mathcal L(B, E)\right] &=&\frac{e^2 B E}{4\pi^2}\left\{ {\rm Li}_2\left(e^{-\pi m^2/(e E)}; e^{-2\pi \frac{B}{E}}\right)+\frac{1}{2} \ln\left(1-e^{-\pi m^2/(e E)}\right)\right\}
    \\
    &=& -\frac{e^2 B E}{8\pi^2} \ln \left[\left(e^{-\pi m^2/(e E)}; e^{-2\pi \frac{B}{E}}\right)_\infty 
    \left(e^{-2\pi \frac{B}{E}}\, e^{-\pi m^2/(e E)}; e^{-2\pi \frac{B}{E}}\right)_\infty
    \right] 
    \\
    &=& \frac{e^2 B E}{8\pi^2}\sum_{k=1}^\infty \frac{\coth\left(\frac{B}{E}\, k\pi\right)}{k} \, e^{-m^2\pi k/(e E)}
    \label{eq:l1be-imag}
\end{eqnarray}
The final expression agrees with the standard result \cite{nikishov,dunne-kogan}, but the second expression is arguably more useful for numerical and asymptotic analysis. From properties of the $q$-Pochhammer function we obtain a different small $q=e^{-2\pi B/E}$ expansion in the weak $E$ field limit
\begin{eqnarray}
    {\rm Im}\left[\mathcal L(B, E)\right] = \frac{e^2 B E}{4\pi^2}\left\{ 
    \sum_{k=1}^\infty e^{-2\pi k\, \frac{B}{E}}\sum_{n|k} \frac{e^{-n \pi m^2/(e E)}}{n} -\frac{1}{2}\ln\left(1-e^{-\pi m^2/(e E)}\right)\right\}
\end{eqnarray}
where the second sum is over the divisors $n$ of $k$.

The real part of the quantum dilogarithm expression \eqref{eq:l1beq-sp11} can be expressed as sums over incomplete gamma functions and exponential integral functions
\begin{eqnarray}
{\rm Re}\left[\mathcal{L}\left(B, E\right)\right] &=&
\frac{e^2 B E}{8\pi^3}\sum_{k=1}^\infty \frac{\coth\left(k \pi\, \frac{E}{B} \right)}{k} \left(e^{i k\pi m^2/(e B)}\Gamma\left(0,\frac{ik\pi m^2}{e B}\right)+e^{-i k\pi m^2/(e B)}\Gamma\left(0, \frac{-i k\pi m^2}{e B}\right)\right)
\nonumber\\
&+&
\frac{e^2 B E}{8\pi^3} \sum_{k=1}^\infty \frac{\coth\left(k \pi\, \frac{B}{E} \right)}{k} \left(e^{k\pi m^2/(e E)}{\rm Ei}\left(-\frac{k\pi m^2}{e E}\right)+e^{-k\pi m^2/(e E)}{\rm Ei}\left(\frac{k\pi m^2}{e E}\right)\right)
\label{eq:l1beq2}
\end{eqnarray}

A similar analysis can be applied to the HE effective lagrangian for scalar QED \cite{euler,weisskopf,dunne-kogan}:
\begin{equation}
    \mathcal L_{\rm scalar}(B, E)=\frac{1}{16\pi^2} \int_0^\infty ds\, e^{-m^2 s}
    \left[\frac{1}{s^3}\left(\frac{(e B s)(e E s)}{\sinh(e B s) \sin(e E s)}-1-\frac{e^2 s^2}{6}(E^2-B^2)\right)\right]
    \label{eq:l1be-sc}
\end{equation}
Invoking the partial fraction expansion \cite{ramanujan}
\begin{eqnarray}
  \frac{1}{s^3}\left(\frac{(e B s)(e E s)}{\sinh(e B s) \sin(e E s)}-\frac{1}{e^2} -\frac{s^2}{6}(E^2-B^2)\right) &=&
  -\frac{2e^2 B E}{\pi} \sum_{k=1}^\infty  \frac{(-1)^k }{k \sinh\left(\frac{E}{B}\, k \pi\right)} 
  \frac{s}{\left(\frac{k\pi}{e B}\right)^2+ s^2}  
  \nonumber\\
  && 
  -\frac{2e^2 B E}{\pi} \sum_{k=1}^\infty  \frac{(-1)^k}{k \sinh\left(\frac{B}{E}\, k \pi\right)} \frac{s}{\left(\frac{k \pi}{e E}\right)^2- s^2}  
\end{eqnarray}
we obtain the following modified Borel integral:
\begin{eqnarray}
  \mathcal L_{\rm scalar}(B, E) &=& -\frac{e^2 B E}{4\pi^3} 
  \int_0^\infty ds\left(
 {\rm Li}_2\left(-e^{-\pi \frac{E}{B}}\, e^{-m^2 s}; e^{-2\pi \frac{E}{B}}\right)
 \left[\frac{s}{\left(\frac{\pi}{e B}\right)^2+ s^2}\right] 
  +{\rm Li}_2\left(-e^{-\pi \frac{B}{E}}\, e^{-m^2 s}; e^{-2\pi \frac{B}{E}}\right) 
  \left[ \frac{s}{\left(\frac{\pi}{e E}\right)^2- s^2}\right]
  \right)
 \nonumber
  \\
  && \hskip -2.5cm  = \frac{e^2 B E}{4\pi^3} 
  \int_0^\infty ds \left(\ln\left[\left(-e^{-\pi \frac{E}{B}}\, e^{-m^2 s}; e^{-2\pi \frac{E}{B}}\right)_\infty\right] \left[\frac{s}{\left(\frac{\pi}{e B}\right)^2+ s^2}\right]
  +\ln\left[\left(-e^{-\pi \frac{B}{E}}\, e^{- m^2 s}; e^{-2\pi \frac{B}{E}}\right)_\infty\right]  \left[ \frac{s}{\left(\frac{\pi}{e E}\right)^2- s^2}\right]\right)
   \label{eq:l1beq-sc20}
\end{eqnarray}
As in the spinor QED case, this formal expression has both a real and imaginary part, with the real part being an integral transform of the imaginary part and also of its modular dual, as a consequence of electromagnetic duality:
\begin{eqnarray}
  \mathcal L_{\rm scalar}(B, E) &=& -\frac{e^2 B E}{4\pi^3} 
  \int_0^\infty ds\,
 {\rm Li}_2\left(-e^{-\pi \frac{E}{B}}\, e^{-m^2 s}; e^{-2\pi \frac{E}{B}}\right)
 \left[\frac{s}{\left(\frac{\pi}{e B}\right)^2+ s^2}\right] 
 \nonumber\\
 &&-\frac{e^2 B E}{4\pi^3} {\mathcal P\mathcal V}\int_0^\infty ds\, 
 {\rm Li}_2\left(-e^{-\pi \frac{B}{E}}\, e^{-m^2 s}; e^{-2\pi \frac{B}{E}}\right) 
  \left[ \frac{s}{\left(\frac{\pi}{e E}\right)^2- s^2}\right]
  \nonumber\\
  &&
  -i\frac{e^2 B E}{8\pi^2} 
 {\rm Li}_2\left(-e^{-\pi \frac{B}{E}}\, e^{-\frac{\pi m^2}{e E}}; e^{-2\pi \frac{B}{E}}\right)
   \label{eq:l1beq-sc2}
\end{eqnarray}
In a ``classical limit'', when $E\to 0$ or $B\to 0$, the expression \eqref{eq:l1beq-sc2} reduces to the correct expressions for $\mathcal L_{\rm scalar}(B)$ and $\mathcal L_{\rm scalar}(E)$, respectively.

The imaginary part of the scalar QED HE effective lagrangian can also be written as
\begin{eqnarray}
    {\rm Im}\left[\mathcal L_{\rm scalar}(B, E)\right] 
    &=& \frac{e^2 B E}{8\pi^2} \ln \left[\left(- e^{-\pi \frac{B}{E}}\, e^{-\pi m^2/(e E)}; e^{-2\pi \frac{B}{E}}\right)_\infty\right] 
    \nonumber\\
 &=&\frac{e^2 B E}{16\pi^2}\sum_{k=1}^\infty \frac{(-1)^{k-1}}{k\, \sinh\left(\frac{B}{E}\, k\pi\right)}\, e^{-m^2\pi k/(e E)}
    \label{eq:l1be-imag-sc}
\end{eqnarray}
The $q$-Pochhammer form leads to a different small $q=e^{-2\pi B/E}$ weak $E$ field expansion:
\begin{eqnarray}
    {\rm Im}\left[\mathcal L_{\rm scalar}(B, E)\right] = \frac{e^2 B E}{8\pi^2}
    \sum_{k=1}^\infty \left(e^{-2\pi k\, \frac{B}{E}}-e^{-\pi k\, \frac{B}{E}}\right)\sum_{n|k} \frac{(-1)^{n}e^{-n \pi m^2/(e E)}}{n} 
    \label{eq:scalar-divisor-sum}
\end{eqnarray}
where the second sum is over the divisors $n$ of $k$.

From \eqref{eq:l1beq-sc2}, the real part of the  HE effective lagrangian for scalar QED can be expressed as sums over incomplete gamma and exponential integral functions
\begin{eqnarray}
{\rm Re}\left[\mathcal{L}_{\rm scalar}\left(B, E\right)\right] &=&
-\frac{e^2 B E}{16\pi^3}
\left\{\sum_{k=1}^\infty \frac{(-1)^k}{k\, \sinh\left(k \pi\, \frac{E}{B} \right)} \left(e^{i k\pi m^2/(e B)}\Gamma\left(0, \frac{ik\pi m^2}{e B}\right)+e^{-i k\pi m^2/(e B)}\Gamma\left(0, \frac{-i k\pi m^2}{e B}\right)\right)\right.
\nonumber\\
&&
\left. 
+\sum_{k=1}^\infty \frac{(-1)^k}{k\, \sinh\left(k \pi\, \frac{B}{E} \right)} \left(e^{-k\pi m^2/(e E)}{\rm Ei}\left(\frac{k\pi m^2}{e E}\right)+e^{k\pi m^2/(e E)}{\rm Ei}\left(\frac{-k\pi m^2}{e E}\right)\right)\right\}
\label{eq:l1beq-sc}
\end{eqnarray}

The scaling relation \eqref{eq:dilog-id1} between the HE effective lagrangian for spinor and scalar QED generalizes to a background field with both magnetic and electric components. The relevant classical dilogarithm identity, ${\rm Li}_2(x)+{\rm Li}_2(-x)=\frac{1}{2} {\rm Li}_2(x^2)$, generalizes to the following {\it quantum} dilogarithm identity
\begin{eqnarray}
 \ln\left[\left(-\sqrt{q}\, x; q\right)_\infty\right]=-
    \ln\left[\frac{(x^2; q^2)_\infty (x; \sqrt{q})_\infty}{(x; q)_\infty (x^2; q)_\infty}\right]
    \label{eq:q-identity}
\end{eqnarray}
This identity implies a scaling relation between the spinor and scalar QED HE effective lagrangians, generalizing relation  \eqref{eq:dilog-id1}:
\begin{eqnarray}
    \mathcal L_{\rm scalar}(B, E)=-\frac{1}{2}\mathcal L_{\rm spinor}(B, E)-2\mathcal L_{\rm spinor}(B/2, E/2)+\mathcal L_{\rm spinor}(B/2, E)+\mathcal L_{\rm spinor}(B, E/2)
    \label{eq:sp-sc}
\end{eqnarray}
This scaling relation \eqref{eq:sp-sc} also follows from the elementary trigonometric identity, generalizing \eqref{eq:classical-identity}:
\begin{eqnarray}
    \frac{1}{\sinh(e B s)\sin(e E s)}&=&  \coth(e B s)\cot(e E s)+ \coth(e B s/2)\cot(e E s/2) \nonumber\\
    &&
    - \coth(e B s/2)\cot(e E s)- \coth(e B s)\cot(e E s/2)
    \label{eq:quantum-identity}
\end{eqnarray}

\section{Heisenberg-Euler Effective Action and the Non-Compact Quantum Dilogarithm}
\label{sec:quantum-non-compact}

To discuss the connection of the HE effective lagrangian to the {\it non-compact} form of the quantum dilogarithm it is more natural to begin with scalar QED. 
Faddeev's definition \cite{faddeev,kirillov,zagier,fock,Kashaev:2011se,Garoufalidis:2021nwy} of the non-compact quantum dilogarithm is
\begin{eqnarray}
    \Phi_b(x)=\exp\left[\int_{\mathbb R+i\epsilon} \frac{dz}{z}\, \frac{e^{-2 i x z}}{4\sinh\left(z\, b\right)\sinh\left(z\, b^{-1}\right) }\right]
    \label{eq:faddeev}
\end{eqnarray}
Compare \eqref{eq:faddeev} with the scalar QED effective lagrangian in \eqref{eq:l1be-sc}.

The integral representation \eqref{eq:faddeev} is defined in the strip $|{\rm Im}(z)|<|{\rm Im}(c_b)|$, where
\begin{eqnarray}
    c_b:=\frac{i}{2}\left(b+\frac{1}{b}\right)
    \label{eq:cb}
\end{eqnarray}
The identification with the compact quantum dilogarithm \eqref{eq:li2} is given by
\begin{eqnarray}
    \Phi_b(x)=\frac{\left(e^{2\pi b(x+c_b)}; q\right)_\infty}{\left(e^{2\pi b^{-1}(x-c_b)}; \tilde{q}\right)_\infty}
    \label{eq:phib}
\end{eqnarray}
where 
\begin{eqnarray}
    q:=e^{2\pi i b^2}\qquad, \qquad \tilde{q}:=e^{-2\pi i b^{-2}}
    \qquad, \qquad {\rm Im}(b^2)>0
    \label{eq:qb}
\end{eqnarray}
In fact, Faddeev's non-compact quantum dilogarithm function \eqref{eq:faddeev} extends to all $b^2$ in the cut plane $b^2\in \mathbb C \setminus(-\infty, 0]$ \cite{faddeev,kirillov,zagier,fock,Kashaev:2011se,Garoufalidis:2021nwy}. 

To relate the non-compact quantum dilogarithm $\Phi_b(x)$ with the HE effective lagrangian for scalar QED \eqref{eq:l1be-sc} we therefore need a phase difference of $\pi/2$ between $b$ and $b^{-1}$. We choose 
\begin{eqnarray}
    b= e^{i\pi/4} \sqrt{\frac{B}{E}} \qquad ; \qquad
    b^{-1}= e^{-i\pi/4} \sqrt{\frac{E}{B}}
    \label{eq:b}
\end{eqnarray}
In the physical Heisenberg-Euler application,  $e B>0$ and $e E>0$, so we are in the regime where ${\rm Im}(b^2)>0$.
This implies that $q$ and its modular dual $\tilde q$ in \eqref{eq:qb} become:
\begin{eqnarray}
    q=e^{-2\pi B/E} \qquad, \qquad \tilde{q}=e^{-2\pi E/B}
    \label{eq:qs}
\end{eqnarray}
We recognize these as the $q$ parameters in \eqref{eq:q1}-\eqref{eq:q2} which appear in the scalar QED HE lagrangian \eqref{eq:l1beq-sc2}. With these identifications, we can write the shift factors in \eqref{eq:phib} in a more natural form for scalar QED:
\begin{eqnarray}
    e^{2\pi b\, c_b}&=& e^{i \pi (b^2+1)}=-e^{-\pi B/E}=-\sqrt{q}
    \\
    e^{-2\pi b^{-1} c_b}&=& e^{-i \pi (1+b^{-2})}=-e^{-\pi E/B}=-\sqrt{\tilde q}
\end{eqnarray}
Therefore, we can express \eqref{eq:phib} as
\begin{eqnarray}
    \Phi_b(x)=\frac{\left(-\sqrt{q}\, e^{2\pi b x}; q\right)_\infty}{\left(-\sqrt{\tilde{q}} \, e^{2\pi b^{-1} x}; \tilde{q}\right)_\infty}
    \label{eq:phib-sc}
\end{eqnarray} 
which should be compared with the HE effective lagrangian for scalar QED in \eqref{eq:l1beq-sc2}.
Rotating the $s$ integral in the first and second terms in 
\eqref{eq:l1beq-sc2} by $e^{-i \pi/4}$ and $e^{+ i \pi/4}$, respectively, we obtain a new expression in terms of $\Phi_b(x)$:
\begin{eqnarray}
    {\mathcal L}_{\rm scalar}(B, E)&=&
    \frac{i e^2 B E}{4\pi^3}\int_0^\infty ds\, \frac{s}{1-i\, s^2} \ln \left[\frac{\left(-\sqrt{q}\, \exp\left(-\frac{\pi\, m^2\, b\, s}{e\sqrt{B E}}\right); q \right)_\infty}{\left(-\sqrt{\tilde q}\, \exp\left(-\frac{\pi\, m^2\, b^{-1}\, s}{e\sqrt{B E}}\right); \tilde q\right)_\infty}\right]
    \nonumber\\
    &=&
    \frac{i e^2 B E}{4\pi^3}\int_0^\infty ds\, \frac{s}{1-i\, s^2} \ln \left[\Phi_b\left(-\frac{m^2 \, s}{2 e \sqrt{B E}}\right)\right]
    \label{eq:scalar-new}
\end{eqnarray}
In this expression \eqref{eq:scalar-new}, $b$ and $b^{-1}$ are given in \eqref{eq:b}, and $q$ and $\tilde q$  in \eqref{eq:qs}. The integrand decays rapidly along the real $s$ axis and the expression \eqref{eq:scalar-new} can be evaluated numerically. It has both a real and imaginary part, which are in numerical agreement with the previous expressions \eqref{eq:l1beq-sc} and \eqref{eq:l1be-imag-sc}, respectively.

The strong-field ($e B, e E\gg m^2$) logarithmic behavior follows from \eqref{eq:scalar-new}, using the fact that
\begin{eqnarray}
\Phi_b(0)= \exp\left[ \frac{i\pi}{24}\left(b^2+b^{-2}\right)\right]
\quad \Rightarrow\quad
    \frac{e^2 B E}{4\pi^3}\ln \Phi_b(0)\sim  \frac{e^2 B E}{4\pi^3} \frac{i\pi}{24}\left(i\frac{B}{E}-i \frac{E}{B}\right)
    =\frac{e^2}{4\pi}\cdot \frac{1}{12\pi}\cdot {\mathcal L}_{\rm Maxwell}
    \label{eq:phib0}
\end{eqnarray}
We recognize these factors as the fine-structure constant, $\alpha=\frac{e^2}{4\pi}$, the scalar QED beta function coefficient, $\frac{1}{12\pi}$, and the Maxwell lagrangian, ${\mathcal L}_{\rm Maxwell}=\frac{1}{2}(E^2-B^2)$.

By a similar argument, the HE effective lagrangian \eqref{eq:l1beq-sp2} for spinor QED can be expressed as
\begin{eqnarray}
    {\mathcal L}_{\rm spinor}(B, E)&=&
    \frac{i e^2 B E}{4\pi^3}\int_0^\infty ds\, \frac{s}{1-i\, s^2} \ln \left[\frac{\left(\exp\left(-\frac{\pi\, m^2\, b\, s}{e\sqrt{B E}}\right); q \right)_\infty
    \left(q\, \exp\left(-\frac{\pi\, m^2\, b\, s}{e\sqrt{B E}}\right); q \right)_\infty}{\left(\exp\left(-\frac{\pi\, m^2\, b^{-1}\, s}{e\sqrt{B E}}\right); \tilde q\right)_\infty 
    \left(\tilde{q}\, \exp\left(-\frac{\pi\, m^2\, b^{-1}\, s}{e\sqrt{B E}}\right); \tilde q\right)_\infty}\right]
    \label{eq:spinor-new}
\end{eqnarray}
where $b$ and $b^{-1}$ are given in \eqref{eq:b}, and $q$ and $\tilde q$  in \eqref{eq:qs}.

\section{Conclusion}
\label{sec:conclusion}

The Heisenberg-Euler one loop effective lagrangian has a natural dispersive integral representation involving the quantum dilogarithm function, for both spinor and scalar QED. The existence of such a dispersion representation was in fact suggested long ago by Lebedev and Ritus \cite{lebedev}. In this paper it is explicitly demonstrated for the one-loop Heisenberg-Euler effective lagrangian. The resulting expressions resum all orders of the real and complex instantons of the Feynman-Schwinger proper-time representation of the effective lagrangian, and they have the form of dispersion integrals relating the real and imaginary parts. The imaginary part is given by the quantum dilogarithm, and the real part is an integral transform of the imaginary part and its modular dual. This is a consequence of electromagnetic duality.
Furthermore, the spinor QED and scalar QED effective lagrangians are related by scaling relations that are encoded in the quasi-periodicity properties of the quantum dilogarithm.

Since the HE effective lagrangian generates all multi-leg photon amplitudes in the constant background field limit, these integral representations are a natural starting point for characterizing the analytic properties of QED amplitudes with many external photon lines. The constant background field limit is indeed a special case, characterized by two independent Lorentz invariant quantities, $(E^2-B^2)$ and $E\cdot B$. This is somewhat analogous to 4-leg scattering amplitudes, characterized by two independent Lorentz invariant Mandelstam variables. Under analytic continuation, QFT scattering amplitudes have interesting properties with direct physical significance, especially when considering more than one Mandelstam variable, transforming between different channels. Here, for the generating function of QED amplitudes in the Heisenberg-Euler limit, we have shown that when considering both Lorentz invariant quantities for the external field the analytic structure becomes much richer, and is directly connected to electromagnetic duality. 

The extension to the effective action for {\it inhomogeneous} background fields corresponds to introducing momenta for the external photon lines, so this becomes an even more interesting problem. In \cite{harris} it has been shown that when the background field depends on an inhomogeneity parameter it is possible to describe the effect of the inhomogeneity on the resurgence properties of the effective action. New Borel singularities are generated, and their nonperturbative effect can be seen in the resurgent analysis of the asymptotic perturbative expansion of the effective action, where the coefficients now depend on the inhomogeneity parameter. It would be interesting to apply the ideas in this paper to the analysis of the dispersive structure of the effective action in inhomogeneous fields. Another interesting extension is to higher loop HE effective lagrangians, building on the seminal 2-loop results of Ritus \cite{ritus} and the partial 3-loop results of Schubert et al \cite{schubert}.

\acknowledgements{I thank Antonino Di Piazza, Holger Gies, Felix Karbstein and Matthew Schwartz for helpful discussions and comments. This material is based upon work supported by the U.S. Department of Energy, Office of Science, Office of High Energy Physics under Award Number DE-SC0010339.}

\end{document}